# Bio-Derived Graphite from Pterocarpus marsupium Leaves for rGO-$MoO_3$ Nanocomposites with Enhanced Photocatalytic Efficiency

*P. Princeya Mary[1], M. Kumaresavanji[2,#], P. Sundara Venkatesh[3], N. Kannan[3] and V. Vasumathi[1,*]*

[1]*PG & Research Department of Physics, Holy Cross College (Autonomous), Affiliated to Bharathidasan University, Tiruchirappalli – 620002, Tamil Nadu, India*

[2]*PG & Research Department of Physics, National College (Autonomous), Affiliated to Bharathidasan University, Tiruchirappalli – 620001, Tamil Nadu, India*

[3]*Nanomaterials Laboratory, Department of Physics, Sri S. Ramasamy Naidu Memorial College, Sattur, 626 203, Tamil Nadu, India*

*Corresponding Authors: #vanji.hplt@gmail.com & *vasumathi@hcctrichy.ac.in

**Abstract**

This study presents a sustainable approach to synthesize bio-graphite from Pterocarpus marsupium (Indian Kino) leaves without using chemical catalysts, activating agents, or organic solvents. The resulting bio-graphite was used to produce rGO via a modified Hummers method. The bio-graphite derived rGO was further incorporated with orthorhombic structured $MoO_3$ at different percentages (1, 3, and 6 wt.%) using ultrasonication. Structural, morphological, and functional characterizations were conducted using XRD, Raman Spectrum, FTIR, FESEM, HRTEM, BET and UV-Vis DRS spectroscopy, revealing a bandgap of 2.82 eV for the rGO(3 wt.%)-$MoO_3$ composite. Photocatalytic activity was evaluated via methylene blue degradation under natural sunlight. The rGO(3 wt.%)-$MoO_3$ nanocomposite showed superior performance, achieving 90% degradation in 150 minutes when compared to 65% by pure $MoO_3$. The Scavenger tests confirmed superoxide radicals ($\bullet O_2^-$) as the main reactive species. This work highlights the potential of bio-graphite derived rGO-$MoO_3$ nanocomposites as efficient, eco-friendly photocatalysts for wastewater treatment.

**Keywords:** Bio-Graphite, reduced Graphene Oxide, $MoO_3$, Nanocomposites, Dye degradation.

## 1. Introduction

Graphene-based materials have emerged as a revolutionary class of nanomaterials due to their exceptional electrical, thermal, mechanical, and optical properties [1,2]. Various methods can be employed to synthesis pristine graphene, graphene oxide (GO), reduced graphene oxide (rGO), and their composites. Recent research highlights that these materials can be produced through sustainable green synthesis methods that encompass plant – derived reduction, microbial or bacterial reduction, and vitamin-assisted reduction [3,4]. Furthermore, the green synthesis method utilizes natural, renewable, and non-toxic precursors derived from agricultural bio-waste materials, such as wood debris, crop residues, and plant-based organic materials [5,6]. Carbon-rich biomass waste, particularly serves as an appropriate precursor for the production of eco-friendly synthetic graphite, which is commonly referred to as bio-graphite. It is non-toxic, biocompatible and suitable for diverse applications due to the absence of heavy metal residues.

These bio-graphitic materials can be converted into graphene oxide (GO) through the introduction of oxygen-containing functional groups, thus enhancing its cost-effectiveness and scalability for large-scale production. Various well – established methods exist for synthesizing GO from graphite by Hummers, Hofmann, Staudenmaier, and Brodie [7–9]. The synthesis entails dispersion of graphite flakes, intercalation of molecules/ions, and finally electrochemical exfoliation. Furthermore, graphene oxide (GO) can be reduced to form reduced graphene oxide (rGO), wherein the removal of oxygen-containing functional groups partially restores the intrinsic thermal and electrical properties of pristine graphene [10]. Owing to its high surface area, excellent conductivity, enhanced charge separation and chemical tunability, rGO continues to serve as a highly efficient material in the photocatalytic application [11-13]. In addition, rGO gains more light harvesting capability in the visible and near-infrared spectral regions, and the presence of residual functional groups makes it more dispersible in various solvents.

In recent decades, semiconductor-based photocatalyst like g-$C_3N_4$, ZnO, $TiO_2$, $Bi_2O_3$, $WO_3$, $Fe_2O_3$, $MoO_3$, CdS, or CuS have been extensively studied due to their UV and visible light activity [14-19]. However, their photocatalytic activity have been constrained due to their restricted surface area, limited absorption of visible-light and rapid recombination of electron-hole pair. To address these challenges, reduced graphene oxide (rGO) has emerged as an excellent scaffold for blending these semiconductor photocatalysts leading to enhance visible light photocatalytic performances [20, 21]

Among these semiconductor-based photocatalyst, orthorhombic Molybdenum trioxide ($MoO_3$) is one of the appropriate n-type semiconductors with a broad optical band gap of approximately 3 eV, which stimulates this material to absorb more UV light [22, 23]. Since $MoO_3$ has more oxygen vacancies, it provides more active sites and readily interacts with water molecules under light irradiation to produce hydroxyl radicals. These radicals play a key role in breaking down the pollutants. However, the photocatalytic activity of $MoO_3$ has been reduced due to its limited electrical conductivity. Its wide bandgap restricts the charge transfer, leading to high recombination of photogenerated carriers. As a result, $MoO_3$ shows low quantum yield, weak reduction capacity, and poor absorption in the visible light region. To resolve this limitation, rGO is used as the best supporting material with $MoO_3$ to reduce the electron-hole recombination and improve its specific functional properties [24]. While previous studies have utilized rGO derived from commercial graphite for compositing with $MoO_3$ [25, 26], the present work focuses on synthesizing rGO from bio-graphite and compositing it with $MoO_3$.

Previous studies have reported the practicability of producing rGO from various bio-waste sources such as coconut shells, rice husks and plant leaves [27-29]. Most of these involve high - temperature treatments and utilization of chemical activators. But, in the present study, we synthesise bio-graphite from Pterocarpus marsupium leaves at low temperature without utilizing any activating agent, chemical catalyst or organic solvent. The Pterocarpus marsupium leaves were chosen as the precursor for the synthesis of bio-graphite due to their rich lignocellulose, which serves as an alternative raw material for the production of carbon material. In addition, the presence of natural polyphenolic compounds such as marsupin, pterosupin, epicatechin, pterostiblene, flavonoids pteroside, pteroisoauroside, carsupin, and liquiritigenin serve as the natural catalyst that facilitates the removal of non-carbon elements during pyrolysis [30]. This eliminates the need for chemical activators, thereby enhancing the graphitization process without introducing any chemical activators.

The resulting bio-graphite from Pterocarpus marsupium serves as the sustainable precursor for rGO, which is subsequently composited with $MoO_3$ in order to establish a sustainable and efficient photocatalytic system. It is expected that the incorporation of bio-graphite derived rGO can enhance the photocatalytic performances of $MoO_3$ under visible light irradiation, which have been analysed and confirmed through various characterization techniques such as XRD, Raman Spectrum, FTIR, FESEM, HRTEM, BET, UV-Vis DRS spectroscopy and photocatalytic analysis.

## 2. Experimental Details

### *2.1. Preparation of bio-Graphite*

Pterocarpus marsupium (Indian Kino Tree) leaves were cleansed and shade-dried for a few days. Then, the dried leaves were pulverized into fine particles and pyrolyzed at a low temperature of about 175ºC, for 3 h. The pre-carbonized sample was mixed with deionized (DI) water and sonicated for 4 h at 90ºC. For every 30 min, the supernatant of the sonicated sample was extracted and replenished with DI water to continue the sonication procedure. The hydro-carbonized sample was then dried at 200ºC for 2 h. Finally, the obtained product underwent a thermal treatment in a muffle furnace at 450°C for 60 min. The synthesized bio-Graphite was characterized by XRD to obtain its phase composition.

*2.2. Synthesis of rGO by modified Hummers' method*

Modified Hummer's method was used to prepare rGO with a few variations in temperature, time length, and oxidizing agent concentration. A certain volume of con. $H_2SO_4$ and 10 g of the obtained bio-graphite were added to a 1000 mL beaker, which was kept in an ice bath. The mixture was allowed to stir for 3 h at a temperature of about 4 – 6ºC. Afterward, $KMnO_4$ was added under controlled conditions to avoid the temperature of the reaction from reaching above 17 °C. The reaction mixture was then stirred vigorously for 40 min, yielding a purple-brown colored suspension. It was then transferred to an oil bath and stirred for 60 min at 45ºC. Following this,75 mL of DI water was added drop by drop to reduce the excess heat production during the dilution process and then stirred continuously for 3 h at 45ºC. Shortly after the addition of 100 mL of $H_2O_2$, gaseous bubbles emerged and the reaction mixture turned dark yellow. The suspension that ensued was washed with aqueous HCl to extract the unwanted residue. Further, the filtrate was ultrasonicated for 20 min at 70°C after being diluted with DI water. This finely dispersed mixture was washed again with DI water and centrifuged until the product reached neutral pH. At last, the prepared product was dried in a hot air oven at 60ºC for 20 h.

*2.3. Hydrothermal synthesis of $MoO_3$ nanoparticles*

The $MoO_3$ nanoparticles are synthesized by adding 0.2 M Ammonium heptamolybdate tetrahydrate $(NH_4)_6Mo_7O_{24}$ to 10 mL DI water under constant stirring. As soon as the reaction mixture was diluted finely, 5 mL of concentrated $HNO_3$ was added in drops. The colloidal solution was then transferred into a Teflon-lined stainless-steel autoclave and heated to 150°C in a hot air oven for 12 h. DI water and ethanol were used to cleanse the resultant product and dried at 70°C for 17 h. Lastly, it was calcined in a muffle furnace at 550ºC for 2 h.

*2.4. Ultrasonic-mediated synthesis of bio–graphite derived rGO-$MoO_3$ nanocomposites*

Using ultrasonic waves, 0.3g of $MoO_3$ was diffused in 50 mL ethanol for 30 min. Following that, a defined ratio of rGO was added to the $MoO_3$ suspension and again ultrasonication was applied for another 30 min. Ultimately, the prepared rGO-$MoO_3$ nanocomposite was washed with DI water and dried at 60ºC for 23 h. Similarly, different percentages (1,3,6 wt.%) of rGO-$MoO_3$ nanocomposites were prepared and labeled as rGO (1 wt.%)-$MoO_3$, rGO (3 wt.%)-$MoO_3$, rGO (6 wt.%)-$MoO_3$.

*2.5. Photocatalytic Degradation Test of Methylene Blue Dye*

Most of the dye industry utilizes Methylene Blue (MB) as a color assiduity, despite its highly poisonous, carcinogenic, and non-biodegradable nature. It also releases a large quantum of processed MB dye into natural water sources, harming both human beings and microorganisms. To degrade the above-mentioned toxic material, photocatalytic performance against MB was performed using bio–graphite derived rGO and $MoO_3$ nanocomposites with three different weight percentages [rGO (1,3,6 wt.%)-$MoO_3$]. For this, a solution of MB dye at a concentration of 10 mg $L^{-1}$ (10 ppm) was prepared. Then, 20 mg of $MoO_3$ and rGO-$MoO_3$ nanocomposites were added separately to each of 50 ml of as-prepared MB solution, and then the photocatalyst-dye mixtures were subjected to dark sonication for 30 min. In dark conditions, the MB dye molecules adhere to the surface of the photocatalyst and it continues until the surface is saturated with MB. The readily adsorbed dye molecule may desorb into the solute in the interim. After some time, the adsorption and desorption reach an equilibrium condition, which will be more helpful in ensuring accurate photocatalytic measurements when irradiated with light. After 30 min in darkness, the photocatalytic test was carried out in direct sunlight on a sunny day in Sattur, Tamil Nadu, India. Every 15 min, 3 mL of the photocatalytic solution was collected, filtered and its adsorption range was measured using a UV-visible spectrophotometer. The maximum absorbance of MB occurred at 664 nm, therefore, the efficacy of the prepared photocatalyst was estimated by monitoring the decrease in the MB's adsorption peak (664nm).

*2.5.1. Reactive Species Quenching Test*

The root cause of the photocatalytic degradation of MB dye is the presence of active species in the photocatalyst. To investigate it, a reactive species quenching test was done utilizing trapping agents such as p-benzoquinone (p-BQ) as superoxide anion quenchers ($\bullet O_2^-$), ethylene diamine tetra acetic acid (EDTA) as hole quenchers, isopropyl alcohol (IP) as hydroxyl quenchers (.OH), and hydrogen peroxide ($H_2O_2$) as electron quenchers. A 0.1 mM of the above scavenger is mixed individually with the optimal photocatalyst [rGO (3 wt.%) - $MoO_3$], and their impact on photodegradation was measured. The scavengers trap their respective radicals, preventing them from reacting with MB dye and resulting in a slower degradation rate. When the degradation rate is low, dominating reactive species can be identified more easily and will be more helpful in framing the photocatalytic mechanism of the highly efficient photocatalyst.

*2.6. Sample Characterizations*

A powder X-ray Diffractometer (PANalyticalX'Pert PRO) equipped with Cu Kα radiation of wavelength λ= 1.5406 Å, an acceleration voltage of 40 kV, and a current of 30 mA was used to determine the crystal structure, phase composition, and crystallite size of the as-prepared samples. Raman spectroscopy was conducted using a LabRAM HR Evolution Laser Raman Microscope (Horiba) with an Oxxius 532nm excitation source (100mw) . The existence of various functional groups was identified using FTIR (Bruker-Alpa, OPUS) spectrometer. A field emission scanning electron microscope (FESEM, Carl Zeiss, Germany) and High resolution transmission electron microscope (HRTEM Thermo Scientific Talos F200S - Schottky FEG, ≤ 0.8 eV, operating at 200 kV with 0.25 nm point resolution) were utilized for the analysis of surface topography and morphology of the material at a very high magnification. Surface area of the samples were analysed using BET instrument (BelshorpMax) and the light absorption properties of the samples had been studied with the help of a UV-visible spectrophotometer (SHIMADZU/UV 2600).

## 3. Results and Discussion

### *3.1. X-Ray diffraction Analysis*

Figure 1 shows the XRD patterns of bio-graphite and rGO, which were synthesized from Pterocarpus marsupium leaves. A prominent diffraction peak is obtained for the as-prepared bio-graphite at a 2θ angle of 29° along the (002) plane. It clearly describes the crystalline nature of graphite and is consistent with JCPDS No. 00-012-0112**.** The observed deviation of the bio-graphite's [002] plane toward a higher degree in the XRD pattern is typically due to the thermal contraction and improved stacking, influenced by the experimental conditions. Specifically, ultrasonication combined with thermal treatment at 90°C for 4 hrs plays the major role in this phenomenon. Effective energy of ultrasonication causes the exfoliation by breaking Van der Waals forces, while the subsequent thermal treatment improves layer restacking and crystallinity. This combined effect initiates the phase transitions , leading to modifications in interlayer spacing and causing the diffraction peak to shift toward higher angles.

As seen in Figure 1(a), peaks at 43° and 47° correspond to the ($100_{[sp1]}$) plane, further supporting the sample's graphitic nature [31]. The less broad peak at an angle of 22.90° implies the presence of less amorphous carbon, which is typically found in graphene oxide or reduced graphene oxide. The Pterocarpus marsupium leaves used in the synthesis of bio-graphite contain inorganic constituents such as calcium, nitrogen, magnesium, and phosphorus [32]. As a result, weak diffraction peaks observed at 35° and 39° correspond to the presence of calcite.

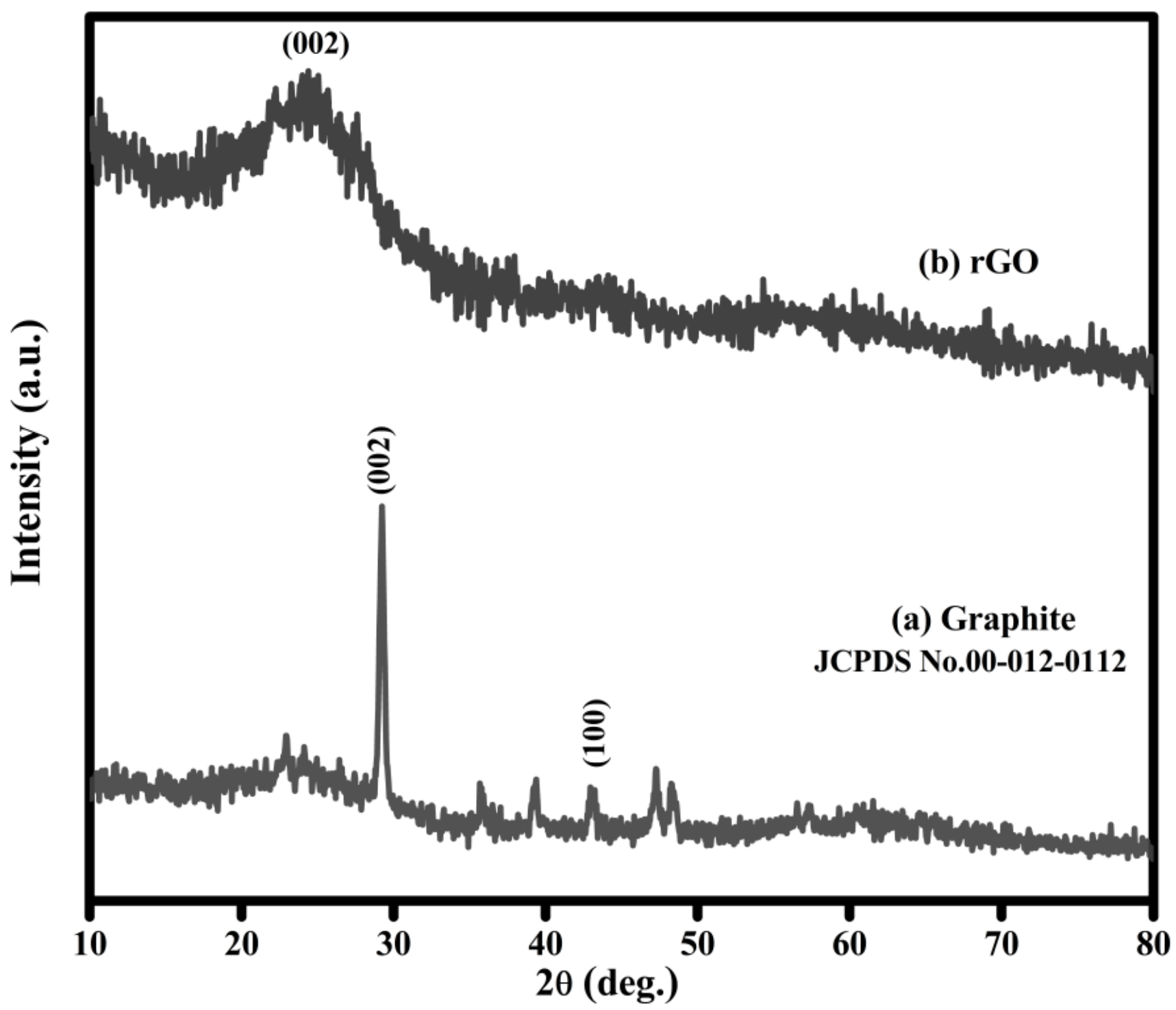

Figure. 1. XRD patterns of (a) bio-graphite and (b) bio-graphite derived rGO.

In addition, the synthesis of bio-graphite was done using DI water without any cleansing agent, which is also responsible for a few elemental impurities along with graphite. Thus, these findings imply that the synthesized bio-graphite engulfs some inorganic impurities due to the natural composition of Pterocarpus marsupium leaves. In the upper XRD pattern, the broad diffraction peak at 23.91° along the crystallographic (002) plane depicts the re-assembly of graphene layers within the formed structure. The computed interplanar spacing of the dominant peak is 0.37nm, which is greater than graphite (0.34 nm), indicating the alteration in graphene layers caused by the reduction process. It also reflects the existence of a lower proportion of oxygen functional groups on the surface and the recovery of the $SP^2$ carbon network; both of these factors are closely tied to the formation of rGO. The formation of rGO instead of GO may be due to the following effects: optimized concentration of Con.$H_2SO_4$ concerning as synthesized bio-graphite, variations in oxidation degree, and adaptation of an ultrasonic-thermal-assisted exfoliation approach during the synthesis process.

The XRD patterns for pure $MoO_3$ and rGO (1,3,6 wt.%) - $MoO_3$ nanocomposites are shown in Figure 2. The observed patterns are consistent with the orthorhombic phase of $MoO_3$, as described in JCPDS No.05-0508 [33]. The sharp and intense diffraction peak along the (040) plane indicates the higher crystallinity, suggesting the well-ordered atoms in the $MoO_3$ lattice. The second dominating peak along the (020) plane indicates the layered structure of $MoO_3$. Thus, the enhanced peak intensities along (040), (020), and (060) planes specify the anisotropic growth of the $MoO_3$ nanostructures. After incorporating rGO with $MoO_3$, the (021) diffraction peak becomes more intense than the (040) peak, indicating that rGO readily aligns and stabilizes the $MoO_3$ layers along the (021) plane, resulting in higher crystallinity.

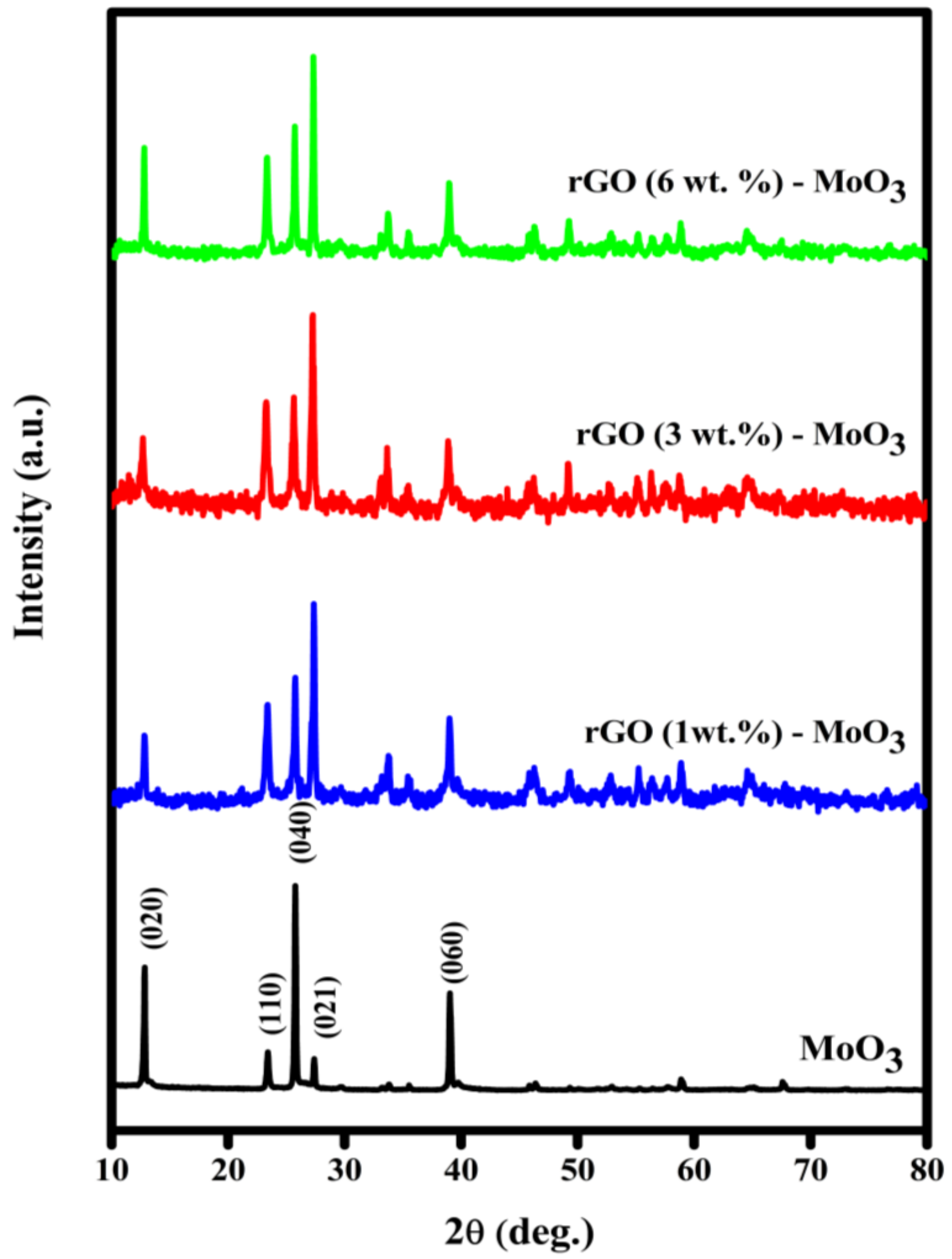

Figure. 2. Powder XRD patterns of $MoO_3$ and rGO-$MoO_3$ nanocomposites.

To assess the structural differences of synthesized composites, the unit cell parameters (a,b and c), crystallite size (D), dislocation density (δ) and unit cell volume (V) are determined using equations (1), (2), (3), and (4), and the obtained results are presented in Table 1.

$$\frac{1}{{d_{hkl}}^2} = \frac{h^2}{a^2} + \frac{k^2}{b^2} + \frac{l^2}{c^2} \quad (1)$$

$$D = \frac{0.9 \times \lambda}{\beta cos\theta} \quad (2)$$

$$\delta = \frac{1}{D^2} \quad (3)$$

$$V = abc \quad (4)$$

Where $d_{hkl}$ is the interplanar spacing between lattice planes of the as-synthesized samples labelled by Miller indices (hkl), a ,b, and c are lattice constants, λ is the wavelength of the X-ray used in the diffractometer, β is the full-width at half maximum and θ is the Bragg diffraction angle.

From Table.1, it is evident that the crystallite size of the $MoO_3$ was reduced from 66.66 nm to 41.38 nm after the addition of 3 wt.% rGO. The observed reduction in crystallite size can be attributed to two primary factors. First, the incorporation of rGO into $MoO_3$ act as a physical barrier during $MoO_3$ crystallization, effectively restricting the growth of larger crystallites. Second, the incorporated rGO occasionally increases the nucleation sites due to the presence of fewer functional groups or defects in it, which results in the formation of smaller crystalites. In contrast, at 6 wt% rGO, the crystallite size increases due to the oversaturation of rGO in the $MoO_3$. The same transition effects had been reflected in dislocation density (**δ)** and unit cell volume (V). These results highlight the importance of optimizing rGO content for enhanced nanocomposite properties.

Table 1. Unit cell parameters, crystallite size, and dislocation density of $MoO_3$ and rGO-$MoO_3$ nanocomposites.

| **Sample** | **a (Å)** | **b (Å)** | **c (Å)** | **Crystallite size D (nm)** | **Dislocation density δ ($nm^{-2}$)** | **Unit Cell volume V (Å)$^3$** |
|---|---|---|---|---|---|---|
| $MoO_3$ | 3.954 | 13.833 | 3.345 | 66.66 | 0.00022 | 182.958 |
| rGO (1 wt.%)-$MoO_3$ | 3.952 | 13.833 | 3.350 | 57.51 | 0.00030 | 183.153 |
| rGO (3wt.%)-$MoO_3$ | 3.972 | 13.891 | 3.362 | 41.38 | 0.00058 | 185.532 |
| rGO (6wt.%)-$MoO_3$ | 3.960 | 13.849 | 3.356 | 62.89 | 0.00025 | 184.101 |

**3.2. Raman Analysis**

Reduced graphene oxide derived from bio-graphite is further confirmed by the Raman spectroscopy. Figure 3 presents the Raman spectra of rGO, featuring two distinct peaks: D band at 1359 $cm^{-1}$ corresponding to disordered carbon and G band at 1588 $cm^{-1}$ indicating the in-plane vibration of graphitic $sp^2$ carbon domain. The sharpness and intensity of G band suggest the restoration of the conjugated graphene network even after the reduction of graphene oxide (GO). The D-band to G-band intensity ratio, used to assess the structural organization, was found to be 0.9. This ratio indicates the moderate defect density that contribute to more amorphous structure as inferred from XRD analysis, along with the transition in hybridization from $Sp^3$ to $Sp^2$ [34]. The persistence of such defects implies that the green reduction method introduces certain functional groups or imperfections that are characteristic of eco-friendly synthesis routes.

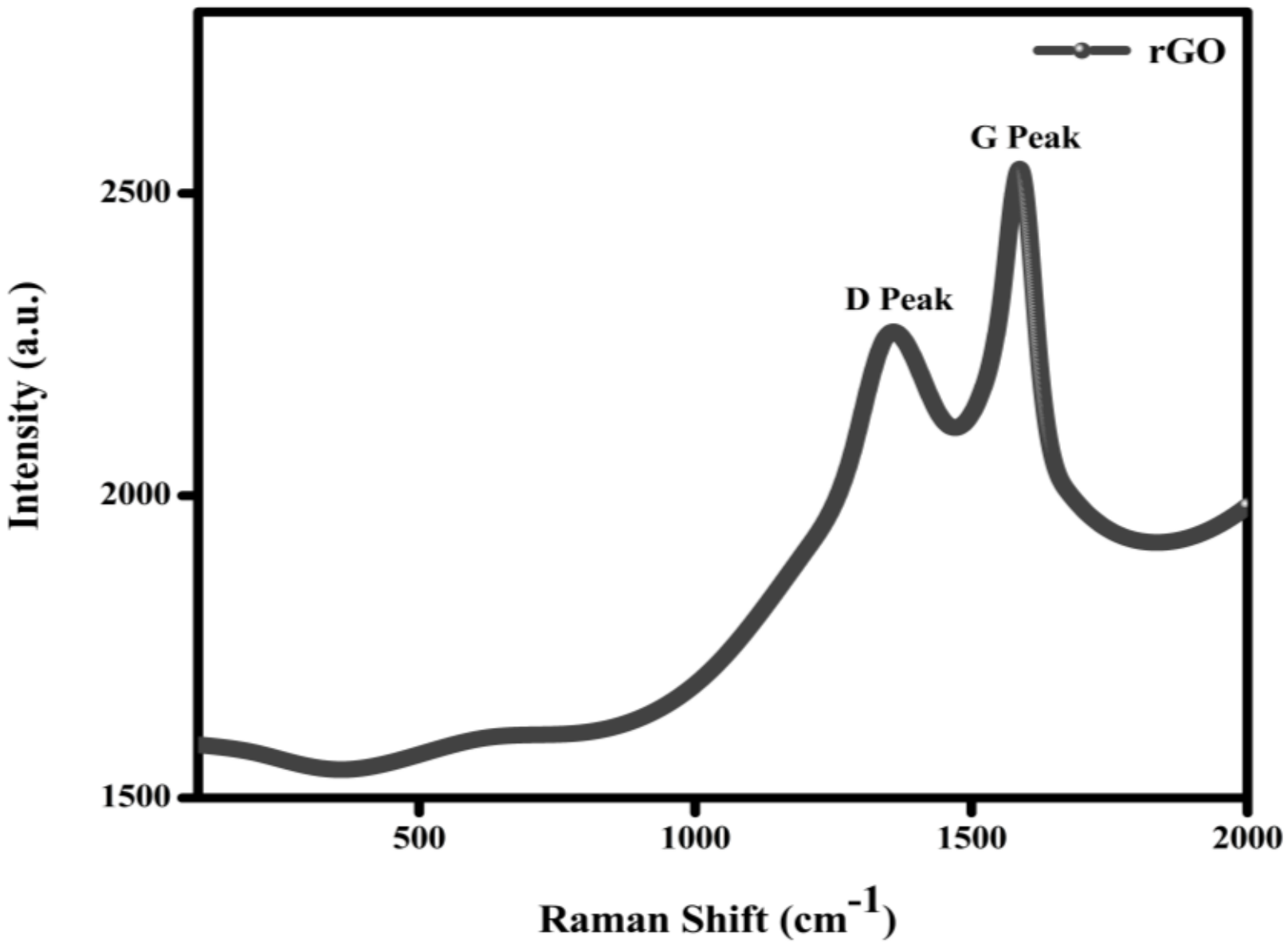

Figure. 3. Raman spectrum of bio-graphite derived rGO.

### *3.3. Functional Group Analysis*

The functional properties of bio-graphite derived rGO and rGO-$MoO_3$ nanocomposites were determined by the FT-IR spectrum in the range between 4000 and 500 $cm^{-1}$ and the results are presented in Figure 4. The spectrum of rGO shows a broad and stretching peak at 3570 $cm^{-1}$ due to the presence of (-OH) hydroxyl groups. It implies that the surface of rGO possesses some oxygen-containing functional groups, which will allow rGO to interact well with water and polar solvents. A subsequent peak at 2739 $cm^{-1}$ reflects the -OH stretching vibration of carboxylic acid, while 2341 $cm^{-1}$ pertains to $CO_2$, most probably due to the adsorption of carbon dioxide on the sample. The band at 1645 $cm^{-1}$ represents C=C stretching vibrations in the aromatic ring, confirming the presence of rGO by conserving the graphene structure [35,36]. In addition, the absorption peak at 650 $cm^{-1}$ caused by out-of-plane C-H bending also defines the characteristics of $sp^2$ hybridized carbon atoms in the graphene structures. At 1231 $cm^{-1}$, the peak illustrates the C-O-C bond stretching, indicative of residual epoxy functionalities in the rGO.

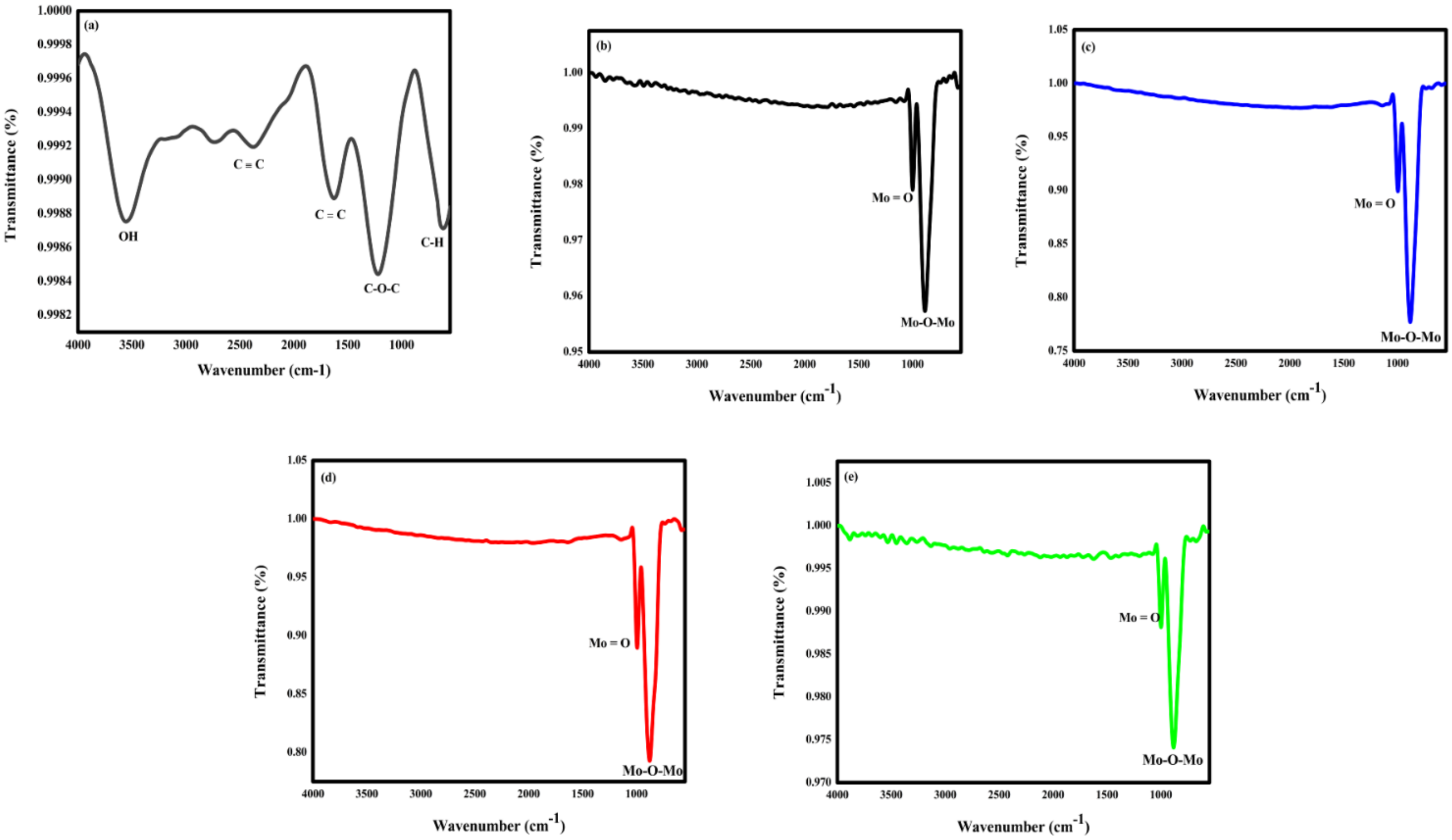

Figure. 4. FT-IR spectra of (a) rGO, (b) $MoO_3$, (c) rGO (1 wt.%) -$MoO_3$, (d) rGO (3 wt.%) -$MoO_3$ and (e) rGO (6 wt.%) -$MoO_3$ nanocomposites

The spectral band of the hydrothermally prepared pure $MoO_3$ shows an asymmetric stretching of Mo-O-Mo bonds at 884 $cm^{-1}$, which implies that Mo atoms are bound by O atoms, forming a three-dimensional network and the distinct vibration of $O_3$ atoms in an orthorhombic crystal lattice. Further, strong absorption at 998 $cm^{-1}$ confirms the existence of a Mo=O stretching vibration. This peak indicates that $MoO_6$ octahedra are linked to form rows, which stack into layers with Mo=O bonds positioned perpendicularly between them. The above two peaks are directly correlated with the XRD results, enabling the conclusion of a well-preserved orthorhombic layered structure [37, 38]. The addition of rGO (1 & 3 wt.%) with $MoO_3$ does not disclose any significant changes. However, some extremely weak peaks in $MoO_3$ have disappeared, leaving just a very weak absorption peak around 1900 $cm^{-1}$ representing a C≡O stretching vibration along with parental peaks. Moreover, the intensity of peaks at 1 wt.% decreases, due to insufficient dispersion, but slightly increases at 3 wt.%, showing a better dispersion. At rGO (6 wt.%), the peak intensity increases significantly and matches the intensity of $MoO_3$, implying a stronger interaction between $MoO_3$ and rGO. As well as it exhibits small characteristic peaks at 3000-3600 $cm^{-1}$ (OH), 2422(C≡O), 1623 $cm^{-1}$ (C=C), and 1463 $cm^{-1}$ (C-O-C), affirming the successful synthesization of rGO-$MoO_3$ nanocomposite.

### *3.4. Morphological Analysis*

### *3.4.1 FESEM Analysis*

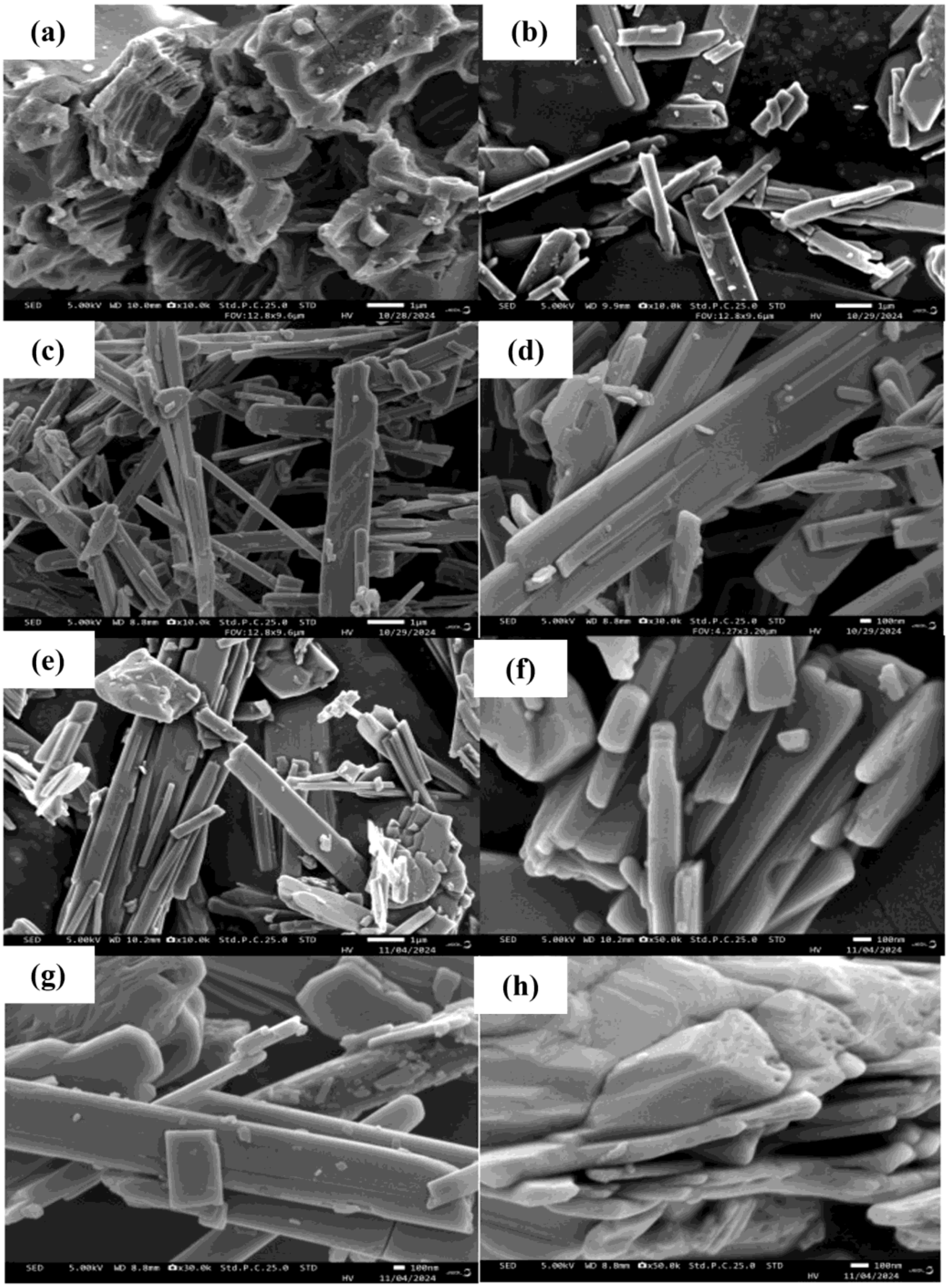

Figure. 5. FESEM image of (a) rGO (b) $MoO_3$ (c-d) rGO (wt.1%) – $MoO_3$ (e-f) rGO (wt.3%) – $MoO_3$ and (g-h) rGO (wt.6%) – $MoO_3$ nanocomposites.

A field emission scanning electron microscope (FE-SEM) was used to evaluate the surface phenomena, such as morphology, and crystallographic information of all the samples . The corresponding micrographs are illustrated in Figure 5. The bio-graphite derived rGO has a typical sheet-like structure with overlapping layers and wrinkles (Figure 5a). Graphene sheets stacked within short range as furrows and ridges, explicitly reveal a substantial reduction in trapped oxygen and other functional groups during the exfoliation process. This observation is further supported by the broad XRD peak at 2θ = 23.91°. Figure 5b portrays the nanostructure of $MoO_3$ that exhibits a belt-like morphology, showing growth along length and width. Its diameter ranges from 0.2-1 ± 0.1 μm and its length is around 0-5 μm. Additionally, it also maintains right angles, which validates the orthorhombic crystal system of $MoO_3$ as correlated with XRD patterns.

In bare $MoO_3$, the layered structure is not visible. However, upon intercalation with rGO, the dispersing ability of rGO improves the visibility of the layered structure, accompanied by an increased number of $MoO_3$ nanobelts. This reflection can be seen in the XRD peak along the (020) plane and the interplanar spacing too. At a concentration of 1 wt.% rGO with $MoO_3$, the layers of nanobelt are well dispersed with a decrease in the diameter of 0.2 -1 ± 0.2 μm and an increase in length of about 0.6 -11 ± 0.6 μm as illustrated in Figure 5c and 5d. At 3 wt.% rGO loading, the layers of $MoO_3$ intersect longitudinally due to the internal wrapping of rGO and it can be seen at some of the edges of the $MoO_3$ layer (Fig. 5e and 5f). Particularly at this concentration, the diameter of $MoO_3$ nanobelt decreases to 0.09-0.92 μm. At the higher concentration of rGO (6 wt.%), the excess rGO aggregates and form thick layers over $MoO_3$ (Fig. 5g and 5h). This in turn blocks the active sites and alters the surface properties, which results in larger crystallite size as well as an increase in the diameter of the belt of around 0.1 - 1 ± 0.8 μm.

### 3.4.2 HRTEM Analysis

The HRTEM analysis of rGO from bio-graphite is shown in the Figure 6 (a) and it exhibits a thin overlapping sheet with varying degree of folding and wrinkles. These characteristics symbolize the presence of multiple graphene layers and highlight the flexible nature of rGO. The edges of rGO sheets appear irregular due to the oxidation condition during the synthesis process. The SAED pattern of rGO in Figure 6 (b) displays the series of concentric diffraction rings, signifying a moderate degree of crystallinity, with partially restored graphene domains are interspersed with amorphous regions. This structural coexistence is due to the defects and residual oxygen functionalities and is align well with the Raman analysis, where the higher intensity D-band indicates the residual defects [39].

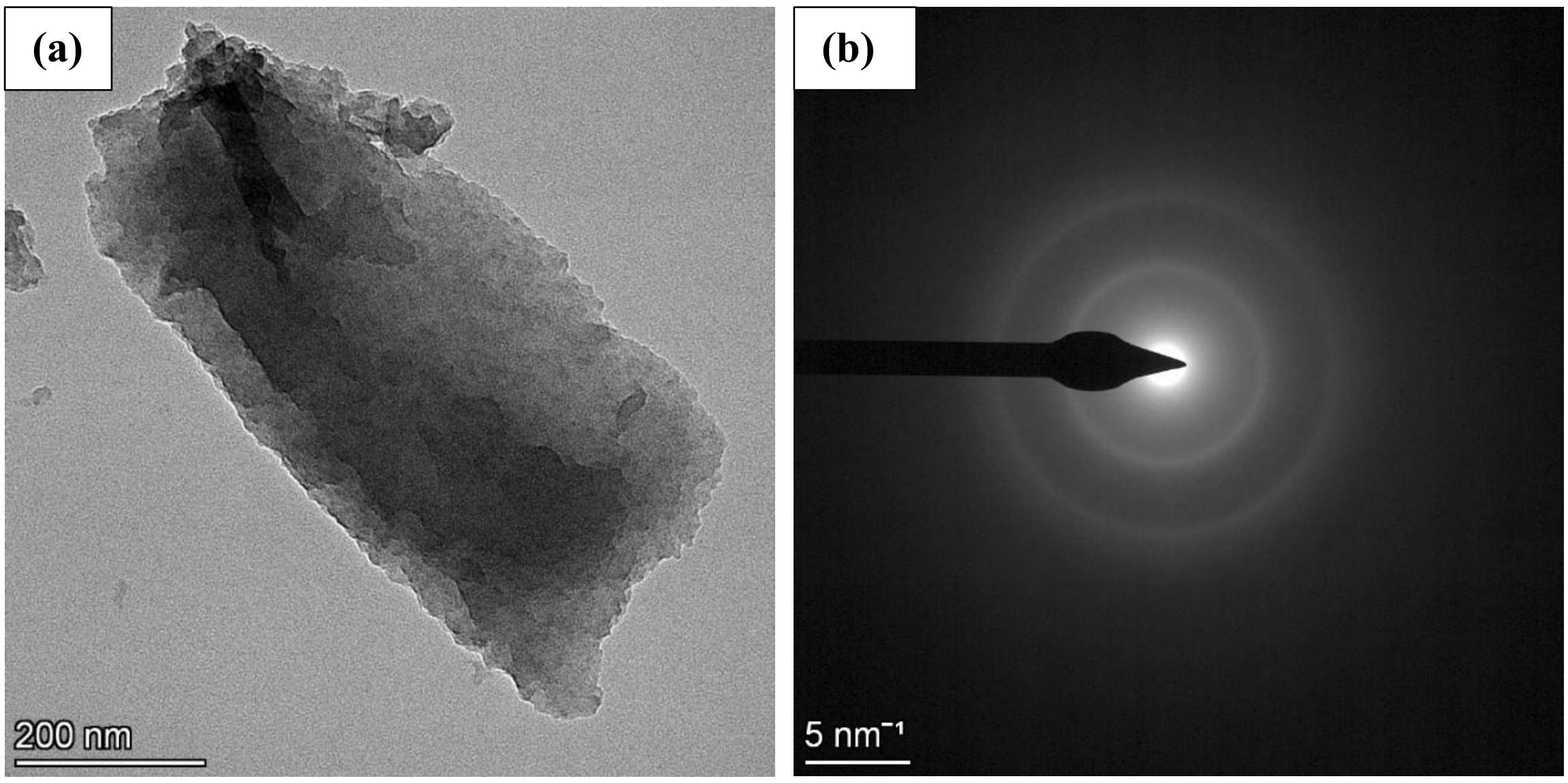

Figure. 6. (a ) HRTEM image and (b) SAED pattern of bio-graphite derived rGO.

**3.5. Brunauer -Emmet-Teller (BET) analysis**

The surface area and the porosity of the $MoO_3$ and rGO (3 wt.%)-$MoO_3$ catalyst were studied using Brunauer -Emmet-Teller (BET) analysis. Initially the photocatalyst sample was degassed at the temperature of 120° C for 3 hrs to remove the moistures and contaminants. Then it was exposed to the adsorbate gas (nitrogen) at a controlled pressure and temperature of 77 K. The adsorption-desorption isotherm was obtained by varying the relative pressure ($P/P_o$) of the adsorbate gas. Figure 7 (a) and (b) show a type-IV isotherm with an $H_3$ hysteresis loop according to IUPAC classification. A sharp increase in adsorption around $P/P_o \approx 0.9$-$1.0$ indicates the occurrence of capillary condensation in slit-shaped mesopores commonly present in layered structure materials. The optimal composite rGO (3 wt.%)-$MoO_3$ shows the similar isotherm pattern but with overlapping of adsorption and desorption branch. It suggests that the incorporated rGO reduces the mesoporosity by filling the pores. The pore size distribution plot of both the catalyst is shown in the Figure 7 (c) and (d). Both $MoO_3$ and rGO (3 wt.%)-$MoO_3$ exhibit the dominant peak around 2nm, which confirms the mesoporous nature of both the samples. After compositing rGO the peak height decreases, indicating a reduction in mesoporous volume as notified in the adsorption -desorption plot [25].

Table 2. Surface area and average pore size of $MoO_3$ and rGO (3 wt.%)-$MoO_3$ nanocomposites

| Sample | Surface Area ($m^2/g$) | Average Pore Size (nm) |
|---|---|---|
| $MoO_3$ | 13.407 | 29.332 |
| rGO (3 wt.%)-$MoO_3$ | 16 | 23.966 |

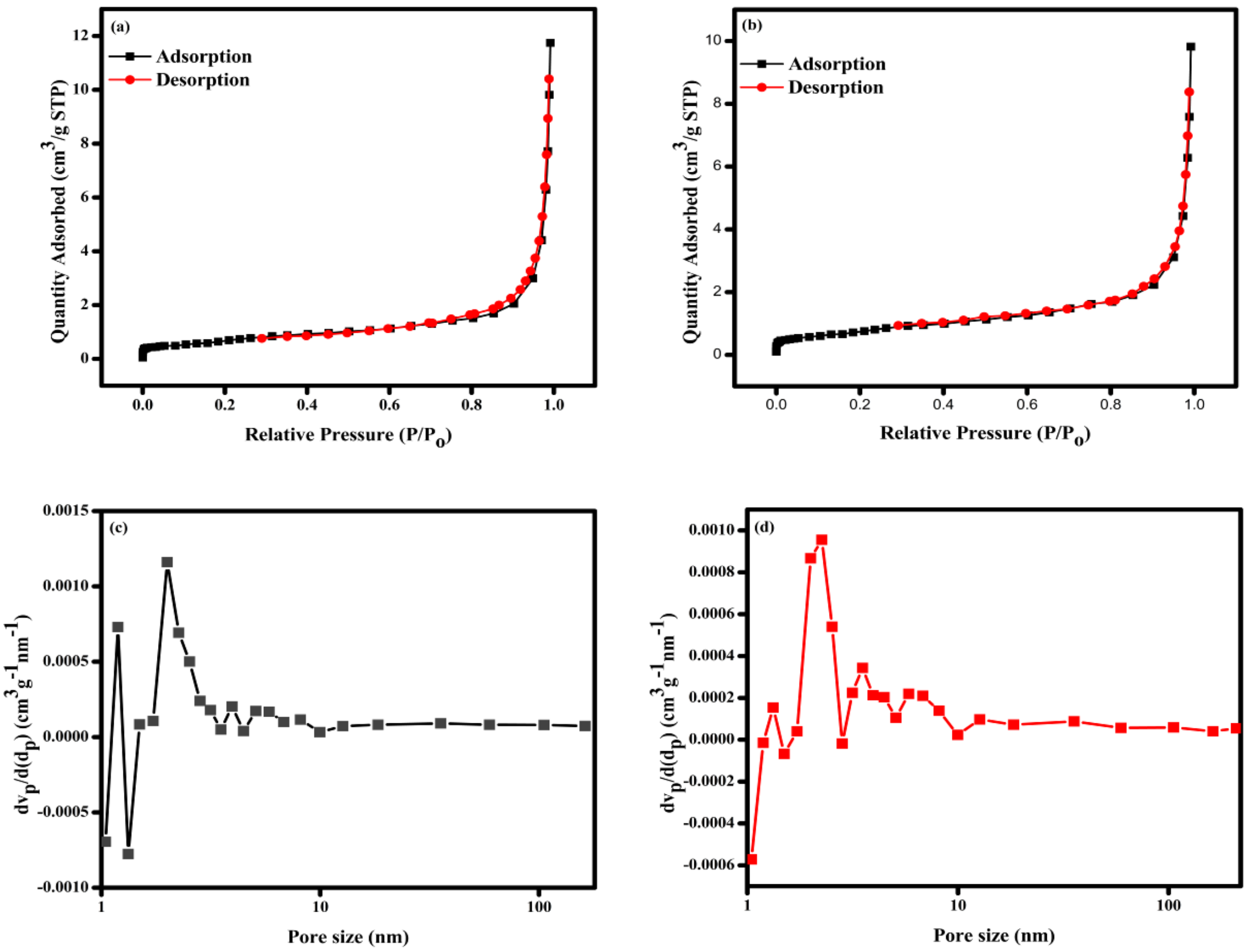

Figure. 7. Adsorption – Desorption isotherms of (a) $MoO_3$, (b) rGO (3 wt.%) - $MoO_3$ and the Pore-size distribution plot of (c) $MoO_3$ (d) rGO (3 wt.%) - $MoO_3$.

The Surface area and average pore size of the $MoO_3$ and rGO (3 wt.%)-$MoO_3$ are presented in Table 2. The orthorhombic $MoO_3$ exhibits a surface area of 13.4 $m^2/g$. Upon compositing the small fraction 3 wt.% of rGO with $MoO_3$, the surface area marginally increased to 16 $m^2/g$. The enhancement after intercalation is modest because the rGO appeared to intercalate between the layers of $MoO_3$, instead of remaining fully exposed, which can be seen in structural morphology of rGO (3 wt.%)-$MoO_3$ (Figure 5 e- f). This internal wrapping reduces the direct contribution of rGO to overall surface area, rather it stabilizes the stacking of $MoO_3$ sheets. This leads to the decreased average pore size, resulting in the formation of uniform mesopores and is further reflected in the pore size distribution plot.

### *3.6. UV-Vis DRS Analysis*

The light absorption properties of all the samples are studied using UV-Visible diffuse reflectance spectroscopy (UV-Vis DRS), and the results are illustrated in Figure 8. As shown in the figure , the reflectance range of all the samples lies within the visible light with different color regions. These values of reflectance are used to calculate the optical band gap of $MoO_3$ and its composites via the Kubelka-Munk function which is expressed as follows.

$$F(R_\infty) = \frac{(1 - R_\infty)^2}{2R_\infty} = \frac{K(\lambda)}{S(\lambda)} \quad (5)$$

Where $R_\infty$, K (λ), and S (λ) denote the diffuse reflectance, absorption coefficient, and the scattering factor, respectively. Figure 9 exhibits the Kubelka-Munk function with respect to photon energy.

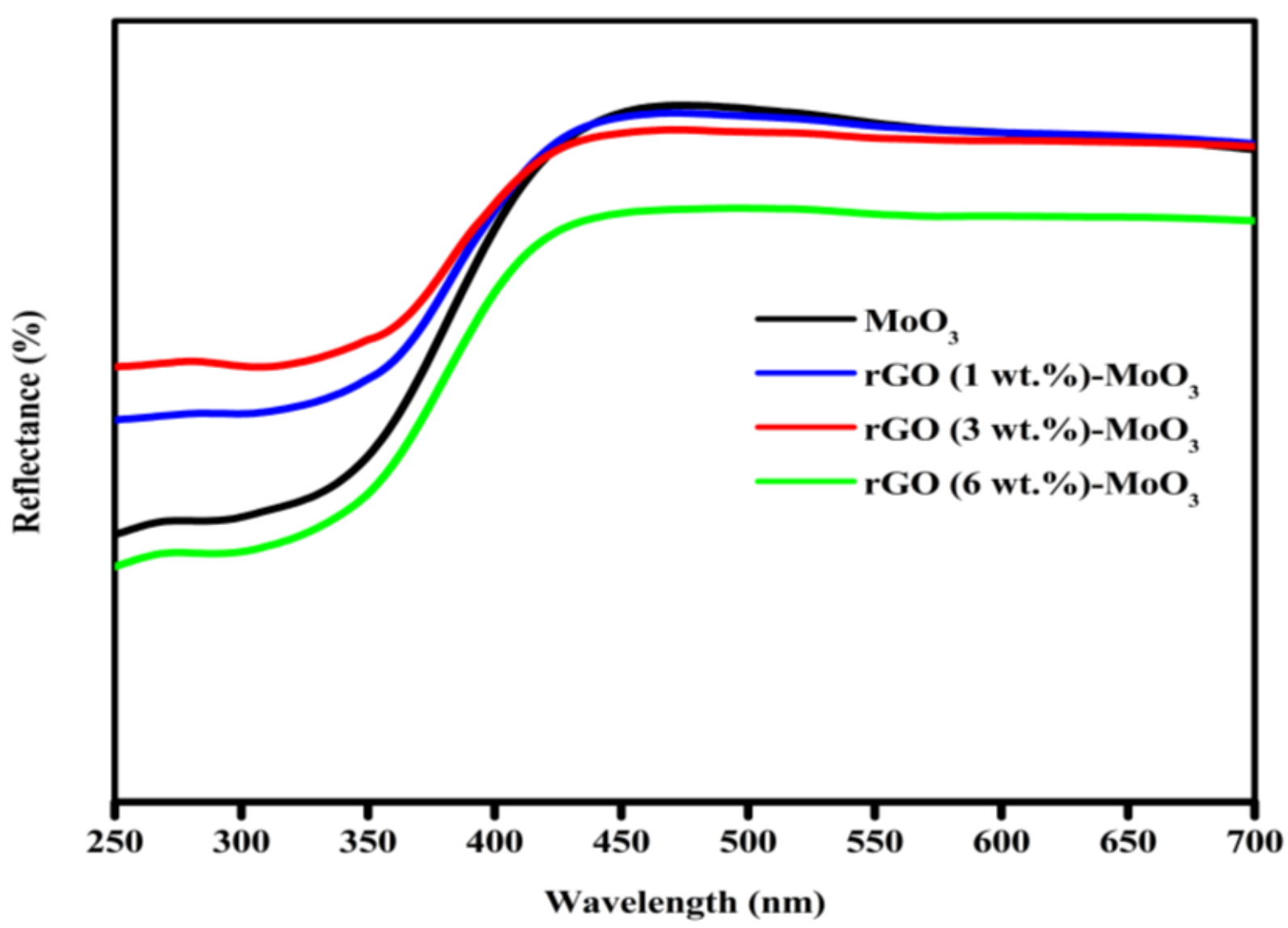

Figure. 8. UV–Vis diffuse reflectance spectra of $MoO_3$ and rGO-$MoO_3$ nanocomposites.

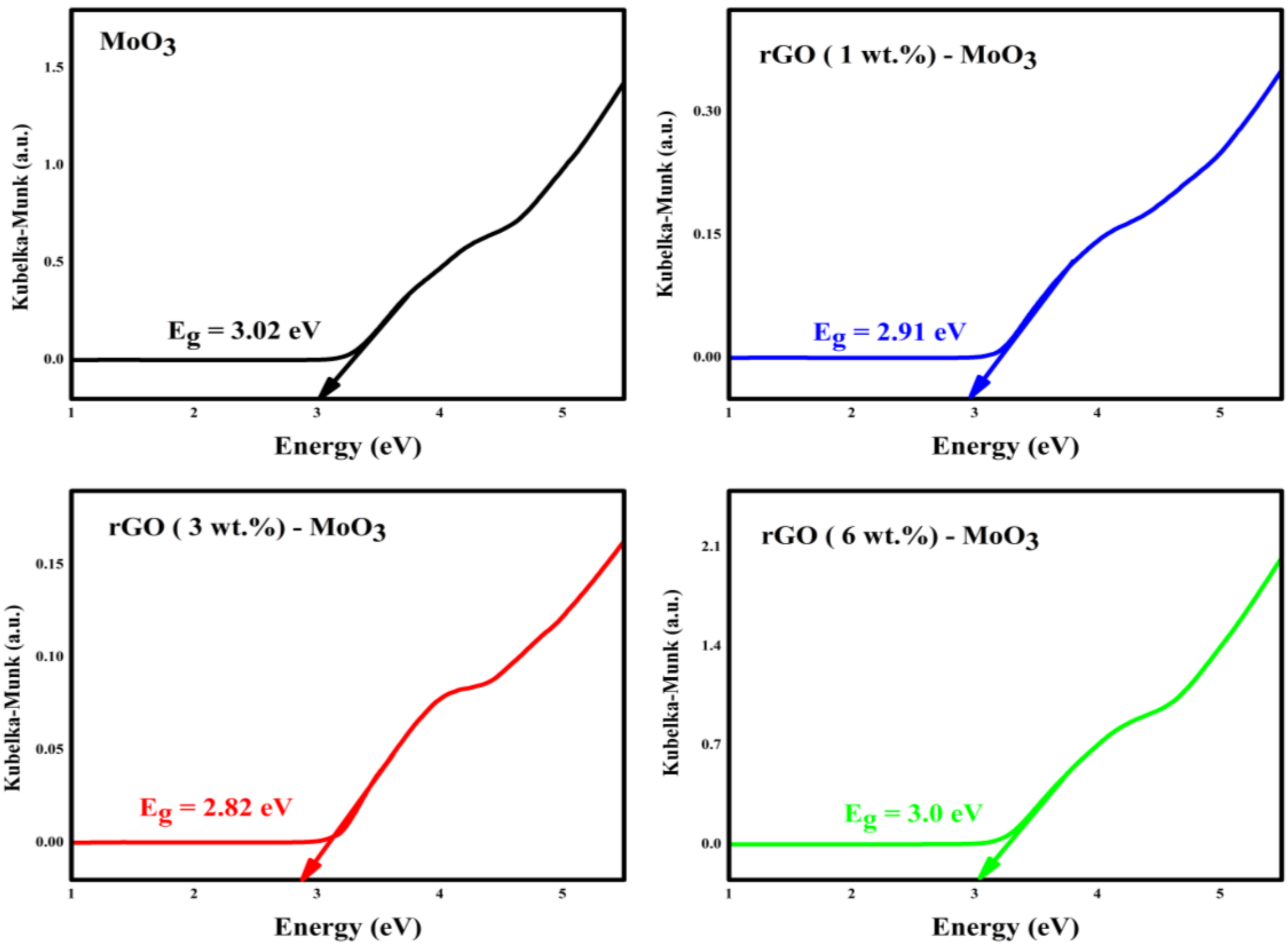

Figure. 9. Kubelka-Munk function vs. photon energy determined from UV-Vis DRS data.

The determined band gap value decreases with an increase in the concentration of rGO till 3 wt.%, which strongly reveals that the optical properties of $MoO_3$ have been enhanced after the incorporation of rGO. The quantum confinement effect of nanomaterials states that the band gap increases with decreasing crystallite size. But in our present work, the band gap energy decreases as the crystallite size decreases, due to the effect of strain caused by the addition of rGO, which is in line with previous results reported by V. Ramar et.al. [40]. The correlation between the band gap and strain values determined from the XRD patterns of $MoO_3$ and its composite is conveyed in Table 3. It is to be noted that band gap energy decreases with increases in the strain value for $MoO_3$, rGO (1 wt.%)-$MoO_3$, and rGO (3 wt.%)-$MoO_3$. For rGO (6 wt.%)–$MoO_3$, the band gap vs strain value shows the opposite effect as we observed in all the other analyses. The higher concentration of rGO tends to clump together or form a layer over the surface, creating blocks or reducing the active sites and resulting in lower efficiency of degradation.

Table 3. Band gap energy and strain value of $MoO_3$ and rGO-$MoO_3$ nanocomposites.

| Sample | $E_g$ (eV) | Strain value (x$10^{-3}$) |
|---|---|---|
| $MoO_3$ | 3.02 | 2.78 |
| rGO(1 wt.%)-$MoO_3$) | 2.91 | 2.81 |
| rGO(3wt.%)-$MoO_3$) | 2.82 | 3.82 |
| rGO(6wt.%)-$MoO_3$) | 3.0 | 3.24 |

### ***3.5.Photocatalytic analysis***

Methylene blue (MB) is a positively charged heteroaromatic thiazine dye that contains nitrogen [41]. It absorbs visible light more effectively at 664 nm, a monomeric (0 – 0) band transition process, and a decreased maximum absorption at 611 nm, which corresponds to a dimeric form. Hence, a distinctive absorption peak of about 664 nm was chosen to track photocatalytic degradation.

Figure 10 (a-d) presents the photodegradation of MB dye and depicts that the maximum absorption at 664 nm decreases for most wavelengths. This declination indicates the gradual breakdown of pollutants over time. Utilizing the absorbance at 664 nm across different time intervals, the percentage of degradation is computed using equation 6 and the results are tabulated in Table 4. The degradation efficiency is determined using the following equation,

$$\text{Degradation Efficiency} = \left[\frac{A_o - A_t}{A_o}\right] * 100 \quad (6)$$

Where $A_o$ is the absorbance of the pollutant at t =0 and $A_t$ is the absorbance at various time intervals, such as t =15, t =30, upto t =150 minutes**.** From the degradation efficiency bar chart [Figure 10 (e)], it is observed that the $MoO_3$ photocatalyst degrades 65.24% of MB dye within 150 minutes. Furthermore, the nanocomposites rGO (1 wt.%)-$MoO_3$, rGO (3 wt.%)-$MoO_3$ and rGO (6 wt.%)-$MoO_3$ achieve 83.76 % , 90.43 % and 75.94 % of degradation over the same duration. Similarly $C/C_o$ vs time plot presented in Figure 10 (f) illustartaes the degradation rate over time, which provides the comparative photocatalytic efficiency of $MoO_3$ and its rGO -incorporates variants. The slowest degradation rate is observed for $MoO_3$, while a notable increase in photocatalytic efficiency occurs as the concentration of rGO increases upto 3 wt.% . However, a slight decline is evident at 6 wt.% of rGO.

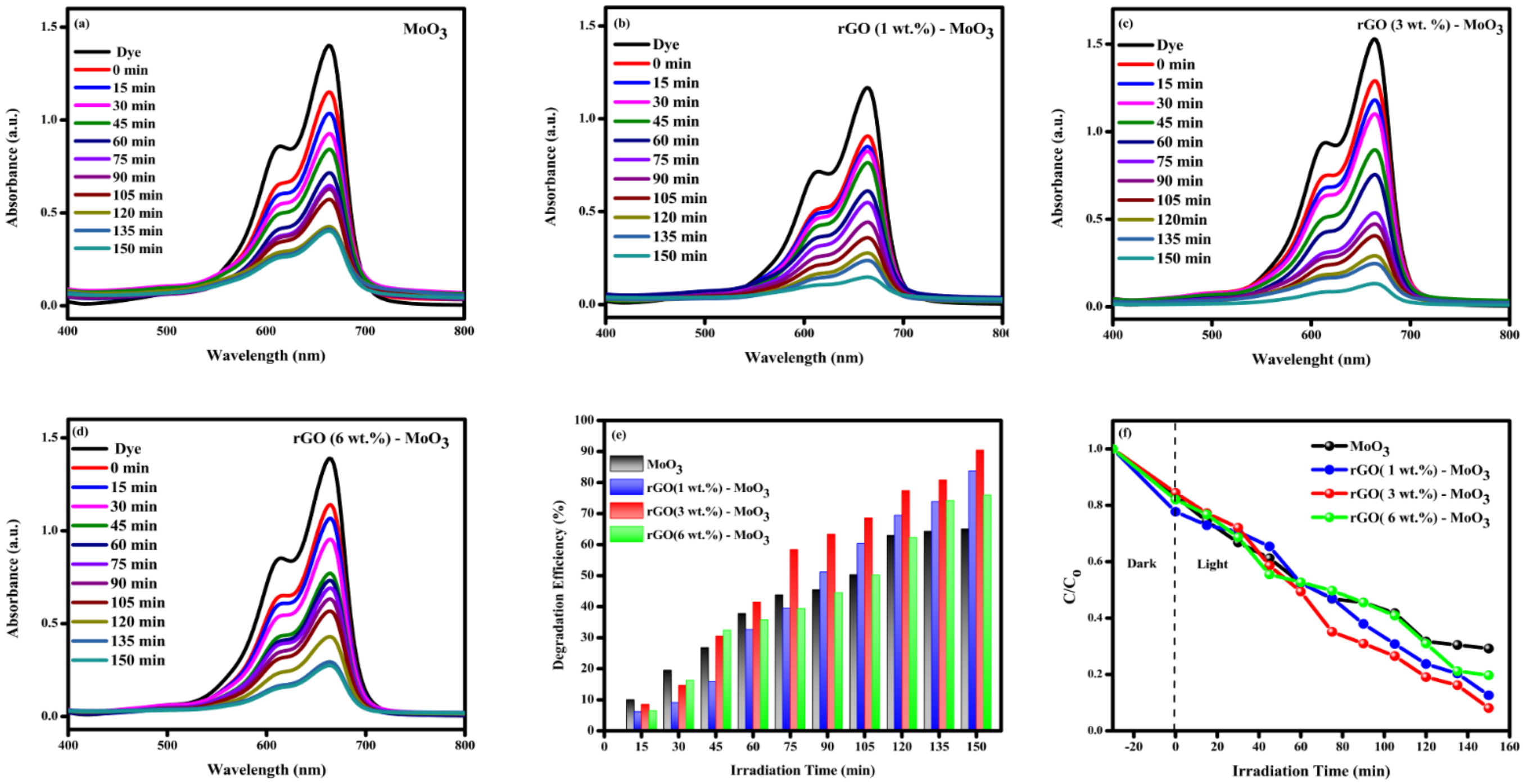

Figure. 10. Photocatalytic degradation of MB dye over (a) $MoO_3$, (b) rGO (1 wt.%)- $MoO_3$, (c) rGO (3 wt.%)- $MoO_3$, (d) rGO(6 wt.%)-$MoO_3$ (e) Degradation efficiency bar chart and (f) $C/C_o$ plot.

It is to be noted that when the incorporation of rGO is increased to 6 wt.%, the degradation efficiency drops (a reverse effect) due to an excessive amount of rGO, slowing the photocatalytic activity. Thus, the enhancement in photocatalytic activity after the incorporation of rGO at 1 and 3 wt.% levels can be explained by two main factors. Firstly, the anchoring of rGO over $MoO_3$ results in a significant reduction in crystallite size, accompanied by enhanced exfoliation and stabilization of $MoO_3$ layers. This optimization increases the surface area-to-volume ratio, providing additional active sites for the photocatalytic reaction. Secondly, the strong interaction between rGO and $MoO_3$ tunes the band gap, thereby delaying the electron-hole recombination under sunlight irradiation. All these observations align with the structural and optical findings from XRD, FESEM, and UV-DRS analyses.

When the sunlight falls on the surface of the bio-graphite derived rGO (3 wt.%)-$MoO_3$ photocatalyst, due to its enhanced light harvesting capacity, more photons are absorbed and more electron-hole pairs are generated [42, 43]. The highly conductive nature and increased surface area of rGO in the composite separate the photogenerated electron-hole pairs and efficiently migrate the electrons from the conduction band of $MoO_3$ to the rGO sheets, leaving the holes in the $MoO_3$'s valence band. These separated electrons and holes took part in the redox reactions. Electrons reduce the oxygen molecules to form reactive oxygen species such as superoxide anions, while holes oxidize water or hydroxyl ions ($.OH^-$) to form hydroxy radicals (OH). As an outcome, the formed reactive species degrade the MB dye into non-toxic organic compounds. Thus, the decreased bandgap of rGO (3 wt.%) -$MoO_3$ facilitates better charge carrier dynamics by increasing the electron-hole separation and migration of electrons. Consequenly the recombination of electron-hole pairs is reduced and leads to better photocatalytic activity of the respective composite. Similar enhanced photodegradation has already been reported in other heterojunction materials, which improve light absorption capacity and the separation efficiency of photogenerated carriers [44]. The rate constant of the heterogeneous photocatalytic degradation was determined from the pseudo-first-order kinetics equation,

$$-ln\left[\frac{C}{C_o}\right] = kt \tag{7}$$

Where *C* represents the concentration of the reactant at a given time (t) and $C_o$ represents the initial concentration of the reactant.

Figure 11 represents the linear fit obtained by plotting the natural logarithm of $C/C_o$ against reaction time (t) . The rate constant (K) was estimated from the slope of the linear fit and has been tabulated in Table 4. All the plots of fitting curves show a linear relationship between $\ln(C/C_0)$ and time, confirming that the degradation of MB follows pseudo-first-order kinetics. The pure $MoO_3$ exhibits the slowest degradation rate, which indicates a lower photocatalytic activity. The rGO(1 wt.%)-$MoO_3$ shows a moderate improvement in photodegradation efficiency compared to pure $MoO_3$, as seen from the slightly steeper slope. However, rGO (3 wt.%)-$MoO_3$ shows the highest photodegradation rate constant among all other samples, suggesting an optimal concentration of rGO enhances photocatalytic performance. Further, rGO (6 wt.%)-$MoO_3$ demonstrates a slight decrease in K value compared to the 3 wt.% sample. This phenomenon may be due to the presence of excess rGO, which act as a shielding layer leading to reduced absorption of visible light. From the obtained results, it is evident that incorporating rGO improves the photocatalytic efficiency of $MoO_3$ by enhancing electron transfer, reducing recombination of electron-hole pairs, and increasing surface area. An optimal rGO content (3 wt.%) achieves the best balance between enhanced electron mobility and light absorption. Beyond this concentration, excess rGO hinders photocatalytic activity.

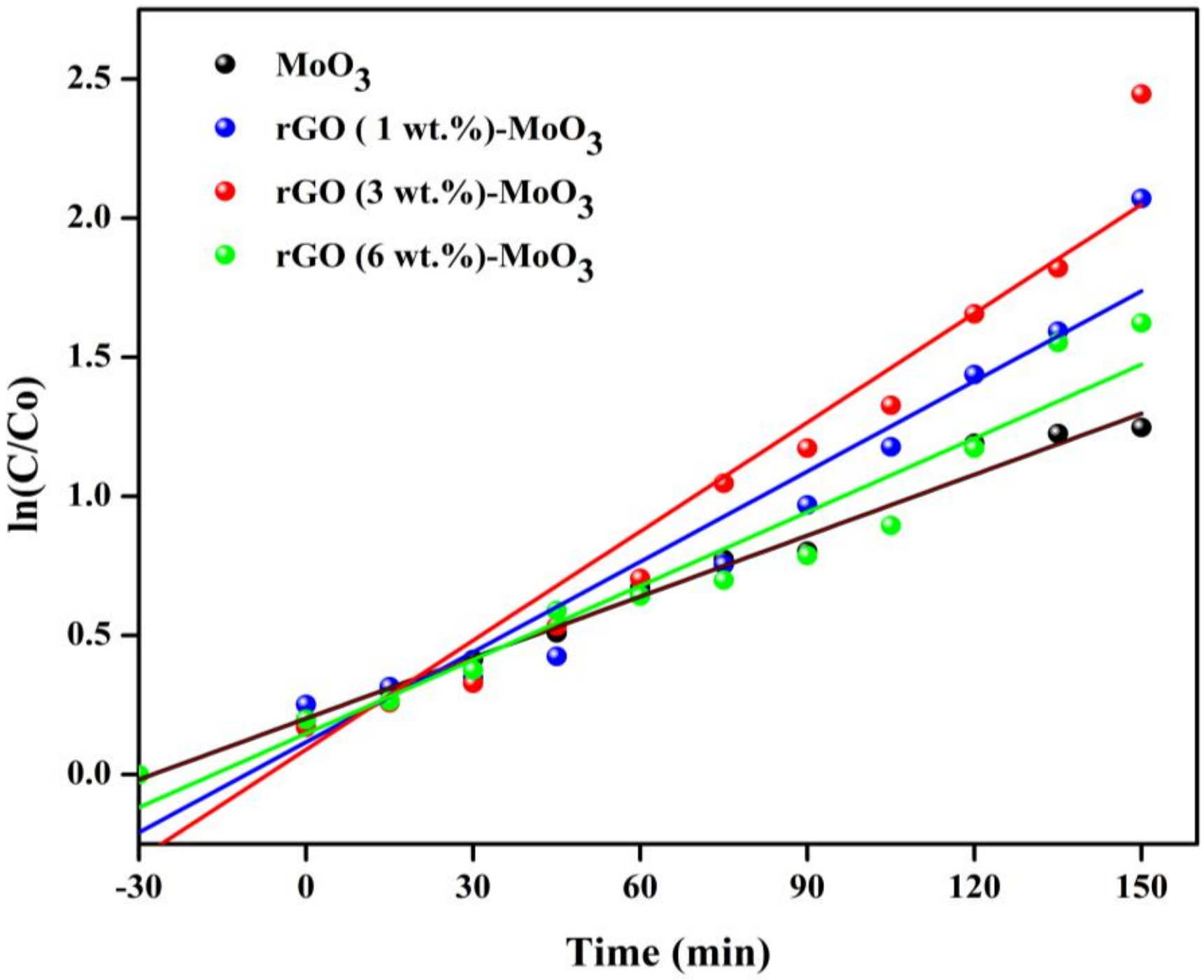

Figure.11. Fitting plots based on a pseudo-first-order kinetics model for MB photodegradation.

Figure 12 illustrates the reactive species quenching test performed over rGO (3 wt.%) –$MoO_3$ nanocomposites. The experimental results reveal the role of the aforementioned radical scavengers in the dye's photodegradation efficiency. The p-BQ predominantly inhibits the degradation efficiency, indicating the significant role of superoxide radicals in the photocatalytic process. Following that, EDTA also shows a comparable degradation, highlighting the contribution of holes in the degradation process. Thus, this study concludes that among all, $\bullet O_2^-$ radical is an active reactive species involved in the degradation of MB, enabled by rGO (3 wt.%)- $MoO_3$ photocatalyst.

The orthorhombic $MoO_3$ has a band gap of 3.02 eV, corresponding to an absorption edge around 413 nm which lies at the boundary between UV and visible regions. As a result, $MoO_3$ primarily absorbs UV light and exhibits reduced absorption in the visible spectrum, thereby limiting the solar-driven photocatalytic activity of $MoO_3$. This defect has been addressed through the incorporation of bio-derived rGO with $MoO_3$. The UV-DRS analysis confirms that the composition of rGO (3 wt.%) with $MoO_3$ tuned the band gap from 3.02 eV to 2.82 eV, significantly enable the ability of composite in harvesting more visible light.

Table 4. Photodegradation efficiency and rate constant of $MoO_3$ and rGO-$MoO_3$ composites.

| Photocatalyst Sample | Photodegradation efficiency (%) | Rate constant K ($min^{-1}$) |
|---|---|---|
| $MoO_3$ | 65.24 | $0.731x10^{-2}$ |
| rGO (1 wt.%)- $MoO_3$ | 83.76 | $1.067x10^{-2}$ |
| rGO (3 wt.%)- $MoO_3$ | 90.43 | $1.246x10^{-2}$ |
| rGO (6 wt.%)-$MoO_3$ | 75.94 | $0.885x10^{-2}$ |

Upon visible light irradiation, the rGO (3 wt.%) –$MoO_3$ photocatalyst generates electron-hole pairs. The rGO, being an efficient electron sink, allows the photogenerated electrons to migrate quickly from the conduction band of $MoO_3$ to the rGO, preventing them from recombining with holes. The transferred electrons in rGO react with oxygen molecules to produce reductive species superoxide radicals ($•O_2^-$), which directly oxidize MB by abstracting the hydrogen atoms, leading to the formation of radical intermediates. Subsequently, it breaks down the pollutant molecules into smaller fragments. Meanwhile, holes in the valence band of $MoO_3$ either directly oxidize dye molecules or react with adsorbed water molecules and generate hydroxyl radicals (•OH) which further degrade the MB dye. Thus, the carried-out scavenger test confirms that superoxide radicals ($•O_2^-$) are the primary reactive species in the photocatalytic degradation of the dye, followed by significant contributions from photogenerated holes ($h^+$). The schematic representation of photodegradation occurred in rGO (3 wt.%) -$MoO_3$ is illustrated in Figure 13. This insight provides a deeper understanding of the reaction mechanism which is also expressed in the following equation form, aiding in the development of optimized photocatalysts for wastewater treatment.

$$MoO_3 + h\nu \rightarrow e^-_{CB} + h^+_{VB} \quad (8)$$

$$e^-_{CB} \rightarrow \text{rGO (3 wt.\%)} \quad (9)$$

$$e^-_{CB(rGO)} + O_2 \rightarrow •O_2^- \quad (10)$$

$$•O_2^- + MB \rightarrow \textit{Mineralization products} \quad (11)$$

$$h^+_{VB(MoO_3)} + H_2O \rightarrow •OH + H^+ \quad (12)$$

$$•OH + MB \rightarrow \textit{Mineralization products} \quad (13)$$

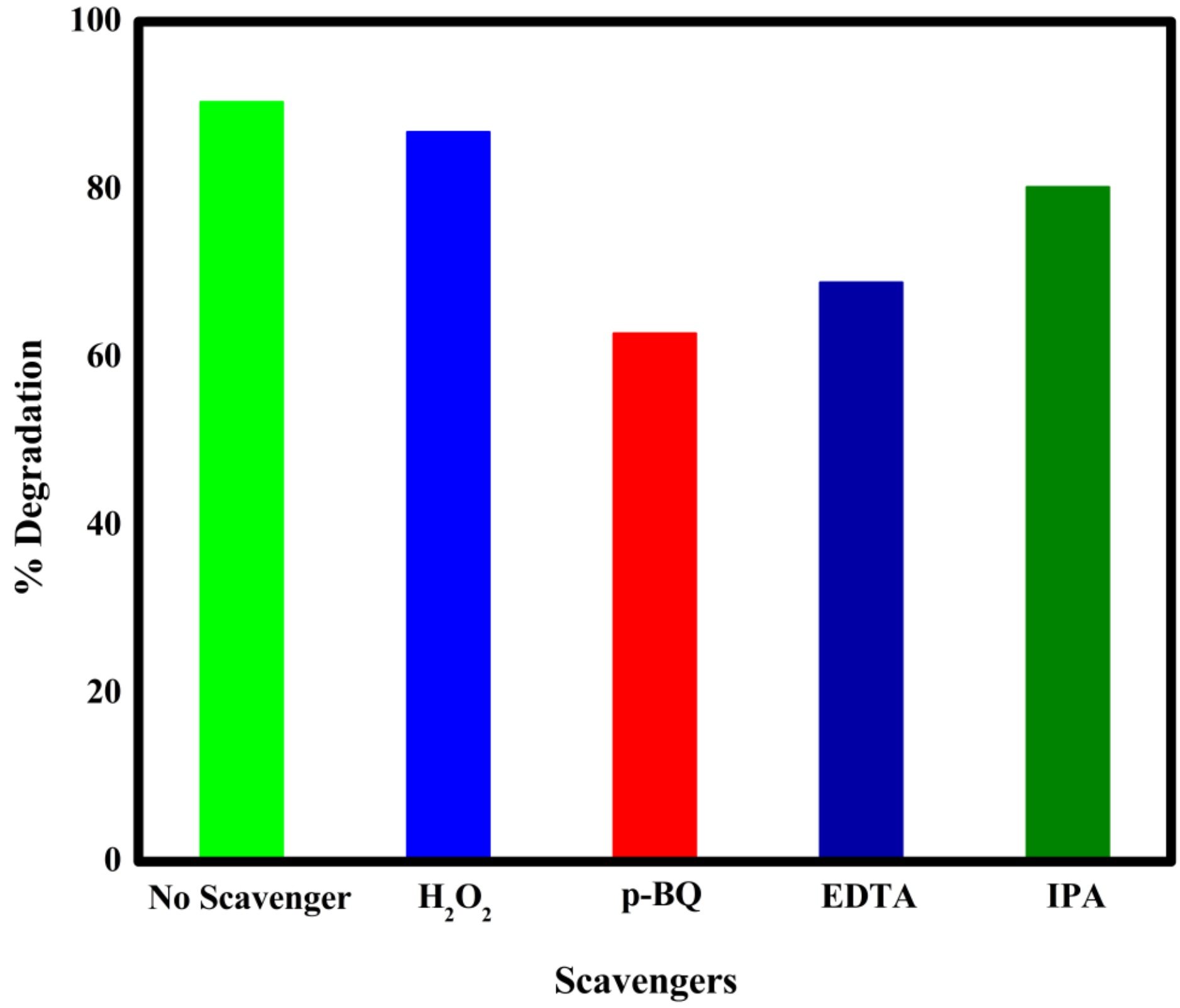

Figure 12. Radical scavengers test on the degradation of rGO (3 wt.%) -$MoO_3$ composites.

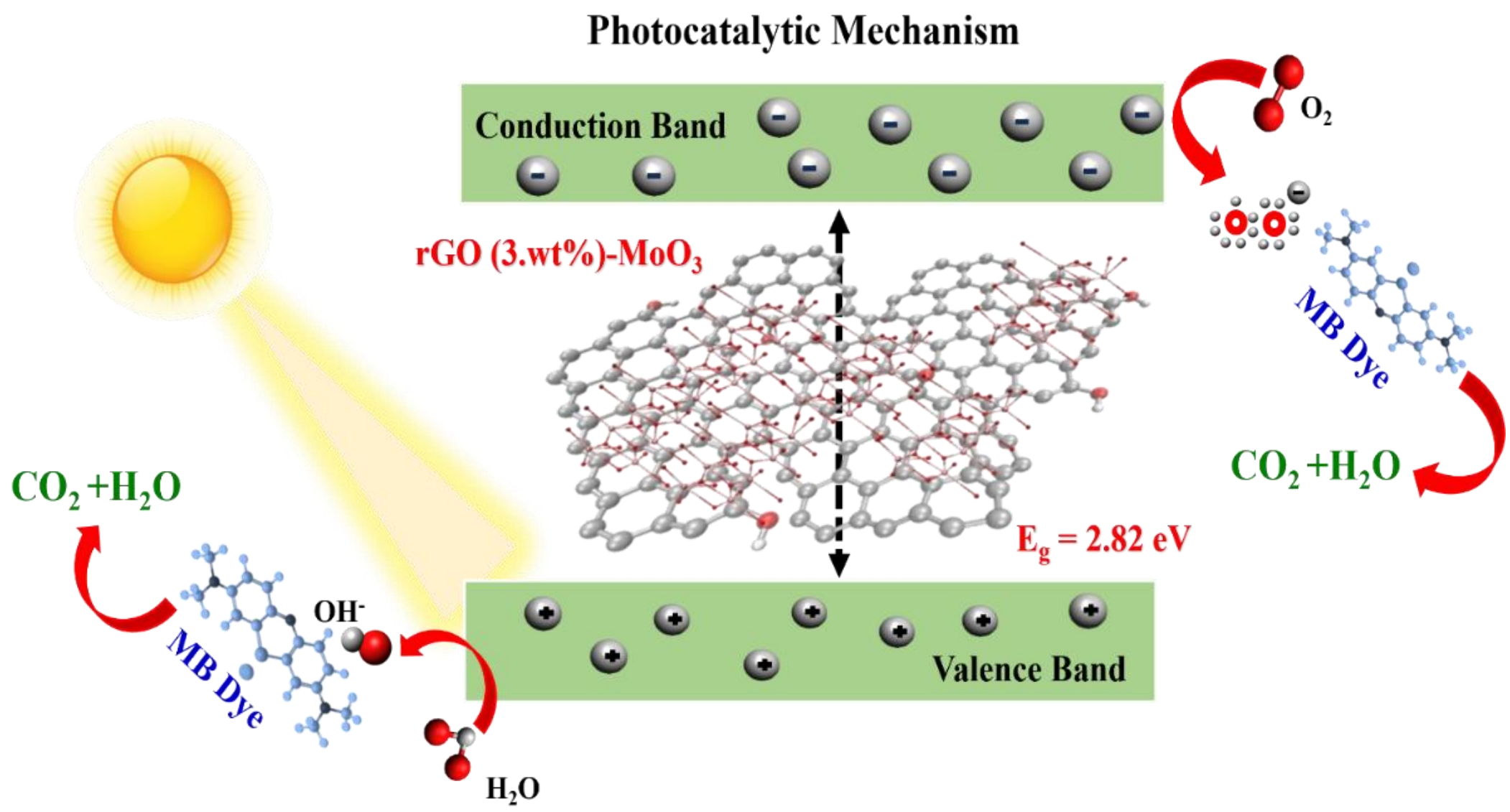

Figure 13. Schematic representation of the photodegradation mechanism in rGO (3 wt.%)-$MoO_3$.

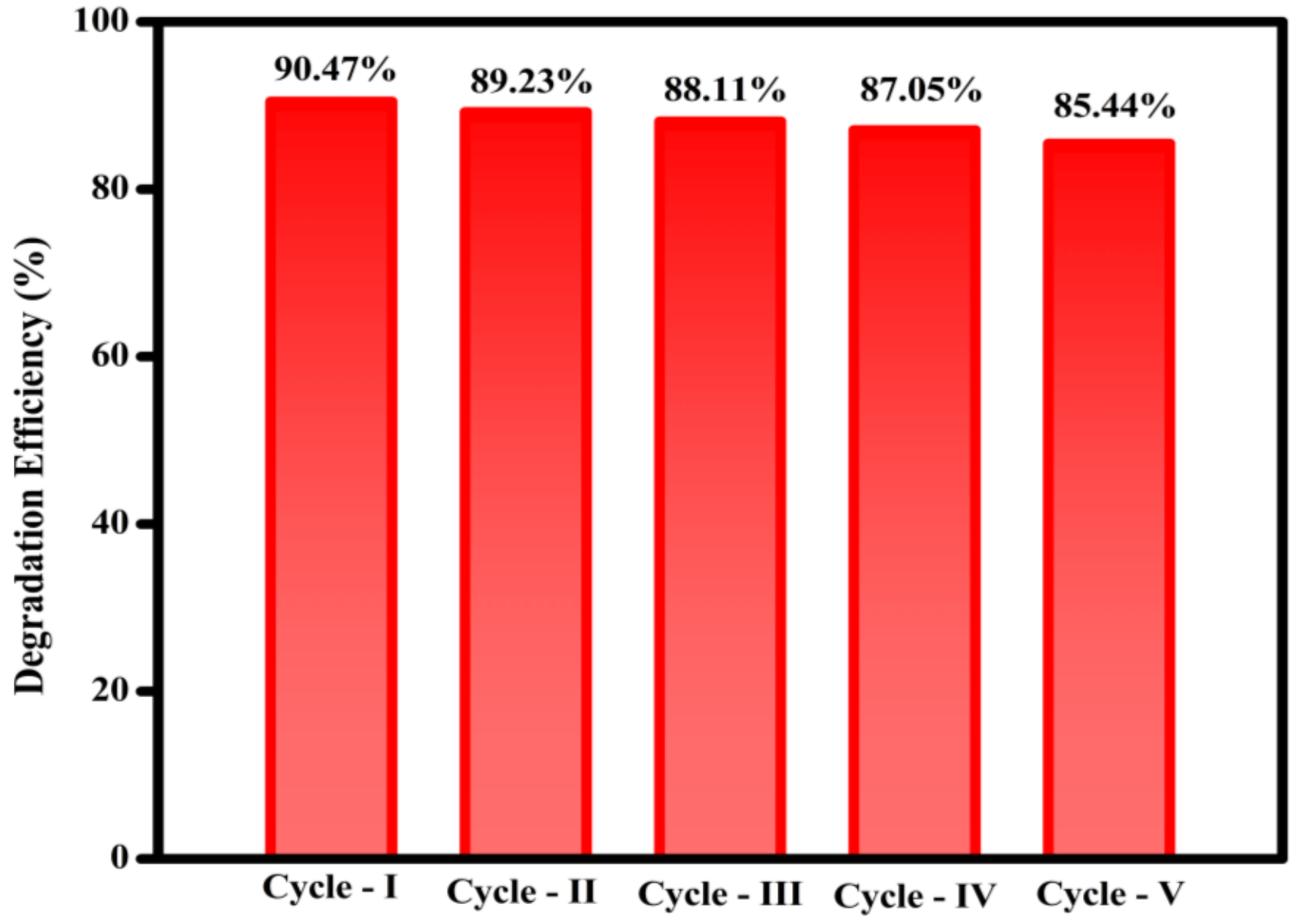

Figure 14. Reusability test of the rGO (3 wt.%)-$MoO_3$ photocatalyst.

To evaluate the stability and reusability of the rGO (3 wt.%)-$MoO_3$ photocatalyst, a series of five consecutive photodegradation cycles were conducted under identical conditions. The degradation efficiency of the photocatalyst was monitored in each cycle which is illustrated in Figure 14. In the first cycle, the nanocomposite demonstrated a high degradation efficiency of 90.47%, indicating excellent photocatalytic activity. In the subsequent cycles, a gradual decline in efficiency was observed. Despite the observed decrease, the rGO (3 wt.%)-$MoO_3$ photocatalyst retained over 85% of its original efficiency after five consecutive uses and highlights its excellent stability and reusability. The observed results indicate that rGO (3 wt.%)-$MoO_3$ is a promising and durable photocatalyst for repeated applications in organic pollutant degradation under visible light irradiation.

Table 5. Comparison of degradation effeciency of rGO (3 wt.%)-$MoO_3$ with previously reported rGO-$MoO_3$ nanocomposites.

| Crystal Structure | Graphite Type | Irradiation Source | Irradiation Time(min) | Degradation Efficiency (%) | Reference |
|---|---|---|---|---|---|
| Orthorhombic | Commercial | UV-light | 120 | 99 | [25] |
| Orthorhombic | Commercial | UV-light | 240 | 50 | [26] |
| Orthorhombic | Bio - graphite (Pterocarpus Marsupium) | Sunlight | 150 | 90 | Present Work |

Further, the photocatalytic degradation efficiency of $MoO_3$ with orthorhombic crystal structure was investigated using different graphite types and irradiation conditions, as summarized in the Table 5. The rGO derived from commercial graphite-based $MoO_3$ photocatalysts exposed to UV-light exhibited varied degradation efficiencies depending on the irradiation time. A high degradation efficiency of 99% was reported after 120 minutes of UV exposure [25]. While another reported only 50% of degradation efficiency after 240 minutes under the similar condition [26]. Conversly, the present study exhibits the high degradation efficiency of 90% under natural sunlight irradiation within 150 minutes, specifically utilizing the rGO derived from bio-graphite (Pterocarpus marsupium) . Thus the bio-derived graphite from Pterocarpus marsupium leaves can serve as an effective, sustainable alternative to commercial graphite in photocatalytic applications. Moreover, the use of sunlight as the irradiation source underscores the potential for cost-effective and environmentally friendly photocatalytic processes.

## 4. Conclusion

This study successfully demonstrates a green route to synthesise bio-graphite from the dried leaves of Pterocarpus marsupium tree using a two-step process involving low-temperature heat treatment and ultrasonication, notably achieved without the use of chemical catalysts, activating agents or organic solvents. The XRD analysis confirmed that bio-graphite tends to form rGO directly with the extra optimized conditions in the modified Hummers' method. Further, the bio-graphite derived rGO was successfully integrated with orthorhombic structured $MoO_3$ at varying weight percentages (1, 3, and 6 wt.%) to fabricate nanocomposites. The resulting composites were characterized for their structural, morphological, surface and optical properties using XRD, Raman, FTIR, FESEM, HRTEM, BET and UV-Vis DRS spectroscopy. The incorporation of bio-graphite derived rGO in $MoO_3$ significantly enhanced the photocatalytic performance under visible light. Among the prepared composites, the bio-graphite derived rGO (3 wt.%)-$MoO_3$ sample exhibited the highest photocatalytic activity, achieving 90% degradation of MB in 150 min. These findings highlight the potential of bio-graphite derived rGO-$MoO_3$ nanocomposites for effective wastewater treatment, offering a cost-effective and sustainable solution for the degradation of organic pollutants.

## CRediT authorship contribution statement

**P. Princeya Mary**: Writing – original draft, Methodology, Investigation, Formal analysis. **M. Kumaresavanji**: Writing – review & editing, Data curation, Validation, Conceptualization, Supervision. **P Sundara Venkatesh**: Data curation, Validation, Investigation. **N. Kannan**: Data curation, Validation. **V. Vasumathi**: Project administration, Supervision.

**Declaration of competing interest**

The authors declare that they have no known competing financial interests or personal relationships that could have appeared to influence the work reported in this article.

**Acknowledgement**

PSV sincerely acknowledges the financial support from the Department of Biotechnology, Ministry of Science and Technology, Government of India, New Delhi, under the STAR College Scheme (HRD-11011/48/2021-HRD-DBT). We acknowledge the Department of Nanoscience and Technology, Bharathiyar University, Coimbatore for extending their BET facility established under the DST-FIST Programme.

**Data availability statement**

All data that support the findings of this study are included in the article.

**Supplementary materials**

Supplementary material associated with this article includes the effect of concentration of photocatalyst on dye degradation efficiency. Figure S1: Impact of concentration of photocatalyst on dye degradation efficiency